\documentstyle[preprint,aps,eqsecnum]{revtex}

\begin{document}
\setcounter{page}{1}


\title{
 Measurement of CPT violation in the neutral kaon system
 \footnotemark[1]
}
\footnotetext[1]
{
 {\it Proceeding for first KEK meeting
on ``CP violation and its origin'' in 1993}
}

\author{
 M. Hayakawa \footnotemark[2] 
 }
\footnotetext[2]
{
Electronic address: hayakawa@theory.kek.jp; \  
JSPS Research Fellow
}

\address{
 Division of Theoretical Physics, KEK, \\
 Tsukuba, Ibaraki 305, Japan
}


\date{\today}

\maketitle

\begin{abstract}

 The basic feature of the measurement
of CPT violation in the neutral kaon system
is overviewed
beginning with noting the physical implication of CPT violation.
 After describing ${\rm K}^0$-$\bar{\rm K}^0$ system at length,
how CPT test is possible is illustrated with adding a few remarks
on it.

\end{abstract}

\pacs{ }

\section{Introduction}
\label{sec:intro}

 The quantum field theory has now become an indispensable tool
on which the fundamental structure of matters
can be described, at least, under Planck scale.
 Once we admit the renormalization prescription,
quantum electrodynamics predicts
the magnitude of the anomalous magnetic moment
of the electron fairly well.
 It is also a great triumph of quantum field theory
that the chiral anomaly,
which is intimately associated with
the regularization procedure in quantum chromodynamics,
explains the decay rate of $\pi^0$ to two $\gamma$s \cite{Jackiw}. 

 The quantum field theory has axiomatic foundation;
relativistic covariance and locality of interaction are
the parts of those axioms.
 The presence of CPT invariance
in any system constructed on such a theoretical foundation
is one of the profound consequences
deduced in the quantum field theory,
where ``CPT'' means the successive application
of C (charge conjugation), P (parity reversal)
and T (time reversal).
 The most important lemma of CPT symmetry
is that the mass and the total decay width
of a particle
must coincide with those of its antiparticle.
 As far as we concern with the calculation of
the cross section or the decay width of some process,
we are not conscious of assurance of those facts
as it is always drawn from the classical Lagrangian
without particular attention.

 Since the present quantum field theory
always assures the CPT invariance,
once the violation of CPT symmetry is discovered in the actual world,
the quantum field theory must be modified
as a foundation describing the elementary structure.
 Thus CPT test is not only the test of one of the discrete symmetries,
but the test of present quantum field theory.

 The possibility of this symmetry violation
is somewhat anticipated
when we recall that the four-dimensional quantum gravity
has not been established as a finite quantum theory yet.
 Of course,
in order for the CPT test to have a practical sense
as a test of quantum field theory,
the magnitude of the remnant of violation
induced from the underlying theory
must be within the reach of the actual measurement.
 This will be the case in the neutral kaon system
as a testing ground of CPT symmetry,
contemplating
that the further precise measurement can be performed,
for instance, by DA$\Phi$NE, FNAL or JHP,
as will be discussed in Sec. \ref{sec:expectation}.

 This report overviews the CPT test in the neutral kaon system
primarily describing the basic subjects concerning with this system
( Sec. \ref{sec:kaon} and Sec. \ref{sec:2pi} )
as well as remarking on the recent result of CPT test at CPLEAR
( Sec. \ref{sec:cpttest} ).
 It does not stick to the content of Ref. \cite{HS}
which was the main ingredient of the talk at the meeting a five years ago,
but taking up the examples
how CPT test has been actually performed
in the experiments in Sec. \ref{sec:cpttest}.
 Sec. \ref{sec:conclusion} is devoted to the conclusion.
 
\section{possibility of CPT violation}
\label{sec:expectation}

 As a possible candidate for yielding the CPT violation
the magnitude of which may be accessible
in the measurement of the neutral kaon system,
we take up the string theory.

 The superstring theory are defined
in the ten-dimensional space-time consistently
as a two-dimensional field theory on the world sheet
\cite{Green}. 
 The CPT invariance has been proved explicitly
in the open superstring theory \cite{Kostelecky}.
 It will be shown in the similar manner
in the framework of the light-cone field theory \cite{Gross}
of heterotic superstring \cite{hetero},
which is the most interesting from the phenomenological side.

 Were the superstring describing the most fundamental structure
of our world,
the dimensional reduction to four-dimension would have to occur
by the dynamics of itself.
 This dynamics is quite non-perturbative.
 As mentioned above, CPT invariance seems to hold
in the present formulation of superstring.
 However the formulation itself is at perturbative sense.
 That is, the CPT invariance has been promised
to hold in the ten-dimensional perturbative level.
 There is no assurance
that the CPT invariant vacuum is realized non-perturbatively,
where CPT is in the four-dimensional sense.

 The various issues from the low energy side
rely solely on the non-perturbative aspect of superstring dynamics;
the origin of masses and weak CP phase,
resolution of strong CP problem,
dynamical breakdown of supersymmetry and electroweak symmetry,
the dimension of space-time.
 If we have no prejudice,
there is nothing to preclude the possibility
of CPT violation induced from such a dynamics.

 Some speculations on the possibility of CPT violation
in string theory can be found in Ref. \cite{CPT-string}.
 The attempt of constructive formulation of superstring
has just begun recently \cite{m-theory},
and the study as quantum theory is
the next issue
including the problem of regularization
and taking proper scaling limit.

 One of the quantity charactering the CPT violation
in the neutral kaon system
is the difference ($M_{11} - M_{22}$)
in the diagonal matrix elements of the Hamiltonian
which describes its time evolution,
the precise definition of which will be given
in Sec. \ref{sec:kaon}.
 Were the higher excitation level states and their interactions
with the standard model particles concern with CPT violation,
its effect on the Hamiltonian would arise with the suppression
by $1/M_{\rm PL}$ most optimistically.
 Then some dimension-two quantity must be accompanied with it
on dimension ground,
which will be the square of kaon mass $m_K$ $\sim 500$ MeV,
the characteristic scale of kaon system.
 Thus we can expect that from the string theory \cite{Ellis}
\begin{eqnarray}
 \displaystyle{
  \frac{(M_{11} - M_{22})_{\rm th}}{m_K}
 }
 &\sim&
  \displaystyle{
   \frac{m_K}{M_{\rm PL}}
  } \nonumber \\
 &\sim&
  \displaystyle{
   0.5 \times 10^{-19}
  }.
\end{eqnarray}

 On the other hand, the recent most bound on the same quantity
due to CPLEAR \cite{CPLEAR}
\begin{equation}
  \frac{(M_{11} - M_{22})_{\rm CPLEAR}}{m_K}
  \lesssim 4 \time 10^{-19},
  \label{eq:CPLEAR-limit}
\end{equation}
which implies that
the existence of CPT violation
may be observed with not so drastic requirement
of improving accuracy,
if it really exists.

\section{Phenomenological description of neutral kaon system}
\label{sec:kaon}

 In order to attach to the measurement of CP violation
as well as to searching the potential existence of CPT violation,
we must start with
describing the time evolution of neutral meson system
as generally as possible admitting several assumptions
such as unitarity of the system.
 It is the most convenient to proceed in such a way that
the unperturbative states of the system
is taken to the eigenstates of the strong and QED interactions,
which will be called as $K^0$ and $\bar{K}^0$
with the degenerate mass $m_K$.
 Those degenerate states are distinguished
from each other by strange number, $S$,
which is conserved in the strong and QED interactions.
 Both of those interactions are recognized as being symmetric under CPT,
CP and T transformations fairly well, and this is assumed hereafter.
 The perturbative part $H_W$
of the Hamiltonian governing the system
contains weak interactions as well as CPT violating sources.
 The time evolution of
the mixed state $\psi$ of $K^0$ and $\bar{K}^0$ is then
given by
\begin{eqnarray}
 &&
 \displaystyle{
  i \frac{d}{dt}
  \left(
   \begin{array}{c}
     c(t) \\
     \bar{c}(t)
   \end{array}
  \right)
 }
 =
 \displaystyle{
  \left(
   \begin{array}{cc}
    M_{11} - i \Gamma_{11}\,/2 & M_{12} - i \Gamma_{12}\,/2 \\
    M_{12}^* - i \Gamma_{12}^*\,/2 & M_{22} - i \Gamma_{22}\,/2
   \end{array} 
  \right)
  \left(
   \begin{array}{c}
     c(t) \\
     \bar{c}(t)
   \end{array}
  \right)
 },
 \label{eq:time}
\end{eqnarray}
where $ c(t) $ and $ \bar{c}(t) $ are defined by
\begin{equation}
 \left| \psi(t) \right>  
   =
   \left(
     \begin{array}{cc}
       \left| K^0 \right>, &  \left| \bar{K}^0\right>
     \end{array}
   \right)
   \left(
     \begin{array}{c}
       c(t) \\
       \bar{c}(t)
     \end{array}
   \right),
\end{equation}
with the proper time $t$ of kaon.

 The connection of the various matrix elements appearing above
with $H_W$ will be found important to observe the properties
under CPT transformation and so on
\footnote{
 The normalization of the states
\begin{equation}
 \left< K({\bf p}) | K({\bf q}) \right> = 
 \delta^3 ({\bf p} - {\bf q}),
 \nonumber
\end{equation}
should be understood,
where $\left| K^0 \right> $ and $\left| \bar{K}^0 \right>$
are orthogonal to each other.
};
\begin{eqnarray}
 M_{11} &=&
 \displaystyle{
  m_K + 
  \left< K^0 \left| H_W \right| K^0 \right>
  + \sum_{n} {\cal P}
    \frac{\left| \left< n \left| H_W
          \right| K^0 \right> \right|^2}{m_K - m_n}
 }, \nonumber \\
 M_{22} &=&
 \displaystyle{
  m_K + 
  \left< \bar{K}^0 \left| H_W \right| \bar{K}^0 \right>
  + \sum_{n} {\cal P}
    \frac{\left| \left< n \left| H_W
          \right| \bar{K}^0 \right> \right|^2}{m_K - m_n}
 }, \nonumber \\
 M_{12} &=&
 \displaystyle{
  \left< K^0 \left| H_W \right| \bar{K}^0 \right>
  + \sum_{n} {\cal P}
    \frac{\left< K^0 \left| H_W \right| n \right>
          \left< n \left| H_W \right| \bar{K}^0 \right>}
         {m_K - m_n}
 },
\end{eqnarray}
where the summation symbol represents
the sum over all possible intermediate states
$n$ involving the integration over momentum configurations.
 These were obtained within the leading order correction in $H_W$.
${\cal P}$ implies taking principal part of the succeeding quantity.
 Likewise the expressions for $\Gamma$'s are
\begin{eqnarray}
 &&
 \displaystyle{
  \Gamma_{11} = 2\pi \sum_{n} \delta(m_n - m_K)
   \left| \left< n \left| H_W \right| K_0 \right> \right|^2
 }, \nonumber \\
 &&
 \displaystyle{
  \Gamma_{22} = 2\pi \sum_{n} \delta(m_n - m_K)
   \left| \left< n \left| H_W
                   \right| \bar{K}_0 \right> \right|^2
 }, \nonumber \\
 &&
 \displaystyle{
  \Gamma_{12} = 2\pi \sum_{n} \delta(m_n - m_K)
   \left< K^0 \left| H_W \right| n \right>
   \left< n \left| H_W \right| \bar{K}^0 \right>
 }.
 \label{eq:gamma}
\end{eqnarray}

 The non-zero off diagonal element causes
the mixing between the unperturbative states,
$ K^0 $ and $ \bar{K}^0 $.
 This implies
that the eigenstates of the time evolution of this system
are not those states in general.
 Such eigenstates are given by the eigenstates
of the 2 $\times$ 2 matrix in Eq. (\ref{eq:time});
\begin{eqnarray}
 &&
 \left| K_S \right> =
 \displaystyle{
   \frac{(1+\epsilon_0+\Delta_{\rm CPT}) \left| K^0 \right>
       + (1-\epsilon_0-\Delta_{\rm CPT}) \left| \bar{K}^0 \right>}
        {\sqrt{
         \left| 1+\epsilon_0+\Delta_{\rm CPT} \right|^2
         + \left| 1-\epsilon_0-\Delta_{\rm CPT} \right|^2}}
 }\,\, , \nonumber \\
 &&
 \left| K_L \right> =
 \displaystyle{
   \frac{(1+\epsilon_0-\Delta_{\rm CPT}) \left| K^0 \right>
       - (1-\epsilon_0+\Delta_{\rm CPT}) \left| \bar{K}^0 \right>}
        {\sqrt{
         \left| 1+\epsilon_0-\Delta_{\rm CPT} \right|^2
         + \left| 1-\epsilon_0+\Delta_{\rm CPT} \right|^2}}
 }\,\, .
 \label{eq:eigen}
\end{eqnarray}
 In the foregoing analysis we take such a phase convention
of CP transformation as
\begin{equation}
 {\cal CP} \left| K^0 \right> = \left| \bar{K}^0 \right> .
 \label{eq:CP}
\end{equation}
 Before writing down
the expression of $\epsilon_0$ and $\Delta_{\rm CPT}$
in terms of the matrix elements $M_{ij}$, $\Gamma_{ij}$,
it is helpful to find the transformation properties
of those elements under CPT and so on.

 For definiteness we would like to see
what CPT invariance implies to $\Gamma_{11}$ and $\Gamma_{22}$.
 Starting to express a part of
the first equation of (\ref{eq:gamma})
in terms of the quantities transformed
by the anti-unitary operator ${\cal CPT}$
\begin{eqnarray}
 &&
 \displaystyle{
  \left| \left< n \left| H_W \right| K^0 \right> \right|^2
 }
 \displaystyle{
  =
  \left| \left< n \left| ({\cal CPT})^\dagger
  \left( {\cal CPT} H_W ({\cal CPT})^\dagger \right)
  {\cal CPT}
  \right| K^0 \right> \right|^2
 }.
\end{eqnarray}
where
$ {\cal CPT} \left| K^0 \right> $ =
$ e^{i \theta_{\rm CPT}} \left| \bar{K}^0 \right> $
in the rest frame of kaons
with arbitrary phase $\theta_{\rm CPT}$
accompanied with T transformation (when recalling (\ref{eq:CP})).
 The particle $n$ will be transformed into its antiparticle
with the opposite direction of momentum.
But as far as the states span the complete basis assuring
the unitarity of the theory (conservation of probability),
it will turn into one of the states already taken into account
under the symbol of summation of the states.
 Thus the expression for $\Gamma_{11}$ in Eq. (\ref{eq:gamma})
becomes
\begin{equation}
 \Gamma_{11} = 2\pi \sum_{n} \delta(m_n - m_n)
               \left| \left< n \left|
                      {\cal CPT} H_W ({\cal CPT})^\dagger
               \right| \bar{K}^0 \right> \right|^2,
 \label{eq:cpt-gamma}
\end{equation}
under CPT transformation.
 Note that CPT invariance has not been assumed until now.
 If we require CPT invariance to the system,
$ {\cal CPT} H_W ({\cal CPT})^\dagger $ = $H_W$.
Then (\ref{eq:cpt-gamma}) becomes nothing but
the expression for $\Gamma_{22}$ in Eq. (\ref{eq:gamma}).
That is, CPT invariance implies that
$\Gamma_{11}$ = $\Gamma_{22}$.
 Similarly one can easily show that
CPT invariance implies that $M_{11}$ = $M_{22}$.
 Those do not depend on the convention (\ref{eq:CP}).
 On the other hand, the statement
that CP invariance implies that both $M_{12}$ and $\Gamma_{12}$
are real is correct under the relation (\ref{eq:CP}).
 We will work in such a convention although
it is possible to proceed in a fashion
independent of any phase convention \cite{phase}.

 From these facts
the difference between $M_{11}$ and $M_{22}$,
for instance, signals the CPT violation,
and the observation of this quantity 
enables us to test CPT invariance
as seen in Sec. \ref{sec:expectation}.

 With understanding that CP, T and CPT violations
is so small that they can be treated as perturbation,
$ \epsilon_0 $ and $\Delta_{\rm CPT}$ in (\ref{eq:eigen})
are also small and are given by
\begin{eqnarray}
 &&
 \displaystyle{
  \epsilon_0 =
  \frac{ -2 {\rm Im}\, M_{12} + i {\rm Im}\, \Gamma_{12} }
       { 2 (\gamma_S - \gamma_L) }
 }, \nonumber \\
 &&
 \displaystyle{
  \Delta_{\rm CPT} =
   \frac{ i (M_{11} - M_{22})
         + \frac{1}{2} (\Gamma_{11} - \Gamma_{22})}
        { 2 (\gamma_S - \gamma_L) }
 },
\end{eqnarray}
where $\gamma_S = i M_S + \Gamma_S /2$
using the mass $M_S$ and the decay width $\Gamma_S$
of the mass eigenstate $K_S$,
and $\gamma_L$ is defined in a similar way.
 They are also easily written down
in terms of $M_{ij}$ and $\Gamma_{ij}$
\begin{eqnarray}
 &&
 \displaystyle{
  m_S = \frac{1}{2}
        \left(
         M_{11} + M_{22} + 2 {\rm Re}\, M_{12}
        \right)
 } , \nonumber \\
 &&
 \displaystyle{
  m_L = \frac{1}{2}
        \left(
         M_{11} + M_{22} - 2 {\rm Re}\, M_{12}
        \right)
 } , \nonumber \\
 &&
 \displaystyle{
  \Gamma_S
    = \frac{1}{2}
        \left(
         \Gamma_{11} + \Gamma_{22} + 2 {\rm Re}\, \Gamma_{12}
        \right)
 } , \nonumber \\
 &&
 \displaystyle{
  \Gamma_L
    = \frac{1}{2}
        \left(
         \Gamma_{11} + \Gamma_{22} - 2 {\rm Re}\, \Gamma_{12}
        \right)
 } .
\end{eqnarray}

 The physical meaning of $\epsilon_0$ and $\Delta_{\rm CPT}$
are clear;
$\epsilon_0$ represents CP and T violations if it is non-zero,
while non-zero $\Delta_{\rm CPT}$ signals CPT and CP violations. 
 Thus $\Delta_{CPT}$ can be said as
a parameter characterizing CPT violation
in the $K^0$-$\bar{K}^0$ mixing.

\section{two-pions decay mode}
\label{sec:2pi}

 In order to confront
with the actual measurement of the neutral kaons,
we have to signify the decay mode appropriate
to the purpose.
 The decaying nature is already evident from
the existence of the ``friction'' terms characterized
by $\gamma$'s appearing in the phenomenological expression
for the time evolution of the neutral kaons.

 The decay to two pions from a kaon
occupies the important status
in the measurement of CP violation as well as CPT violation.
 When we look up one decay channel,
there would arise another source of CP or CPT violation,
which is involved partially in the mixing matrix elements
of kaons.
 This section is devoted to clarifying this aspect for
two $\pi$'s decay
which may help to understand
the principle of CPT test performed at CPLEAR
in Sec. \ref{sec:cpttest}.

 The experiment apparatus detects the particles
each of which has a definite electromagnetic charge
as well as mass.
 For the two-poins states there are two possibility
consisting of such definite charged states;
$\pi^+\pi^-$ and $\pi^0\pi^0$.
 Thus it is tempted to define the parameters
\begin{eqnarray}
 \eta_{+-}
 &=&
 \displaystyle{
  \frac{\left< \pi^+ \pi^- \right| T \left| K_L \right>}
       {\left< \pi^+ \pi^- \right| T \left| K_S \right>}
 }, \nonumber \\
\eta_{00}
 &=&
 \displaystyle{
  \frac{\left< \pi^0 \pi^0 \right| T \left| K_L \right>}
       {\left< \pi^0 \pi^0 \right| T \left| K_S \right>}
 },
\end{eqnarray}
where $K_S$ and $K_L$ are the mass eigenstates
given by Eq. (\ref{eq:eigen}).

 The magnitude of $\eta_{+-}$ and $\eta_{00}$
are expected to be small by the following reason.
 From (\ref{eq:eigen}) and the convention (\ref{eq:CP}),
$K_S$ is almost CP even state whereas $K_L$ almost CP odd.
 On the other hand, as the total angular momentum
of the two-pions resulting from the decay of kaon is zero,
the relative angular momentum of two pions must be zero,
which implies that two-pions state is parity even.
 Hence $\pi^+\pi^-$ and $\pi^0\pi^0$ are both CP even states
in the rest frame of two pions.
 Thus $K_S$ can decay into the two-pions state
without being suppressed by any small CP violating parameters while
$K_L$ can only decay into two pions through the small component,
characterized by the CP violating weight.

 Note that the conservation of angular momentum
is assumed in the above argument.
 Although Lorentz invariance ensures it,
we are left
with the possibility
that Lorentz invariance and CPT invariance is violated
but angular momentum is conserved.
 Of course we may have to reconsider
to relax this assumption. 

 The CP, T or CPT violating components
in the decay amplitude
are not transparent in the physical basis used in the definition
of $\eta_{+-}$ and $\eta_{00}$.
 This originates from the fact
that the charge conjugation operation ${\cal C}$
does not commute with the electromagnetic charge.
( Rather they anti-commute with each other. )
 We compromise with the basis denoted by $K^0$ and $\bar{K}^0$
as for the kaon,
but take the basis consisting of the states invariant
under CP, T or CPT transformation
as possible as for two-pions states.
 This is the case of isospin eigenstates
since the magnitude of isospin is invariant
under CP, T and CPT transformation.

 The two-pions states decompose into
two isospin eigenstates $ I = 0, 2 $
from Bose statistics.
 Explicitly
\begin{eqnarray}
 \left| \pi^+ \pi^- \right>
 &&
 \displaystyle{
  =
  \sqrt{\frac{1}{3}} \left| (2\pi)_2 \right>
  +
  \sqrt{\frac{2}{3}} \left| (2\pi)_0 \right>
 }, \nonumber \\
 \left| \pi^0 \pi^0 \right>
 &&
 \displaystyle{
  =
  \sqrt{\frac{2}{3}} \left| (2\pi)_2 \right>
  -
  \sqrt{\frac{1}{3}} \left| (2\pi)_0 \right>
 }.
 \label{eq:iso}
\end{eqnarray}
 At the stage of examining the transformation property
under CP, T and CPT,
it will be found as convenient to take
the following parametrization;
\begin{eqnarray}
 \left< (2\pi)_I \right| T \left| K^0 \right>
 &=&
 \displaystyle{
  A_I e^{i\,\delta_I}
 }, \nonumber \\
 \left< (2\pi)_I \right| T \left| \bar{K}^0 \right>
 &=&
 \displaystyle{
  \bar{A}_I e^{i\,\delta_I}
 },
\end{eqnarray}
for $ I = 0, 2 $.
 The $\delta_I$ is the phase shift of
S-wave (assuming the angular momentum conservation)
component of $\pi\pi$ scattering
at the invariant mass $ \sqrt{s} = M_K $.

 With these preparations,
we can see the CPT requirement on the decay amplitudes.
 Writing the amplitude using the inner product
\begin{equation}
 A_I e^{i\,\delta_I}
 =
 \left(
   \left| (\pi\pi)_I^{\rm out} \right>,\,
   T\,\left| K^0 \right>
 \right),
\end{equation}
then inserting CPT operation gives
\begin{equation}
 A_I e^{i\,\delta_I}
 =
 \left(
  {\cal CPT} \left| (\pi\pi)_I^{\rm out} \right>,\,
  \left( {\cal CPT}\,T ({\cal CPT})^\dagger \right)
  {\cal CPT} \left| K^0 \right>
 \right)^*,
 \label{eq:inner}
\end{equation}
where the superscript attached on $(\pi\pi)$ signifies
which of the asymptotic states correspond to
the given states,
and the complex conjugation comes
from the anti-unitary nature of CPT operation.
 In the rest frame of two pions we have
\begin{eqnarray}
 \displaystyle{
  {\cal CPT} \left| (\pi\pi)_I^{\rm out} \right>
 }
 &=&
 \displaystyle{
  \left| (\pi\pi)_I^{\rm in} \right>
 } \nonumber \\
 &=&
 \displaystyle{
   e^{2 i \delta_I} \left| (\pi\pi)_I^{\rm out} \right>
 },
 \label{eq:in-out}
\end{eqnarray}
from the definition
$ \left< (\pi\pi)_I^{\rm out} | (\pi\pi)_I^{\rm in}\right> $
= $ e^{2i\delta_I} $.
 Inserting (\ref{eq:in-out}) into (\ref{eq:inner}) leads
\begin{equation}
 A_I e^{i \delta_I}
 = e^{2i\delta_I}
   \left\{
     \left< (\pi\pi)_I^{\rm out} \right|
     \left(
      ({\cal CPT}) T ({\cal CPT})^\dagger
     \right)
     \left| \bar{K}^0 \right>
   \right\}^* .
\end{equation}
 If CPT invariance holds, the quantity in the above bracket
turns into $\bar{A}_I e^{i \delta_I}$ so that
\begin{eqnarray}
 A_I = \bar{A}_I^*,
\end{eqnarray}
from CPT invariance.

 Thus the combination
\begin{equation}
 \frac{\bar{A}_I - A_I^*}{\bar{A}_I + A_I^*},
\end{equation}
represents the CPT violation in the amplitude
of the two-pions decay channel.
 Hereafter we take such a phase convention \cite{Wu}
of the amplitude that 
\begin{equation}
 \frac{\bar{A}_{I=0}}{A_{I=0}},
\end{equation}
is real.
 Then the above quantity for $I=0$ reduces to
\begin{equation}
 \lambda_0 =  \frac{\bar{A}_0 - A_0}{\bar{A}_0 + A_0}.
\end{equation}
 One can express $\eta_{+-}$ and $\eta_{00}$, or
\begin{eqnarray}
 &&
 \displaystyle{
  \epsilon = \frac{1}{3} \left( 2 \eta_{+-} + \eta_{00} \right)
 }, \nonumber \\
 &&
 \displaystyle{
  \epsilon^\prime =
   \frac{1}{3} \left( \eta_{+-} - \eta_{00} \right)
 },
\end{eqnarray}
using the relation (\ref{eq:iso}),
the explicit form of which can be found in
Appendix C of Ref. \cite{HS}.
 Note that $\epsilon^\prime$ is proportional to $\omega$
which is defined by
\begin{equation}
 \omega = \frac{{\rm Re}\, A_2}{{\rm Re}\, A_0}.
\end{equation}
 Numerically $1/\omega = 20 $,
the example of the empirical fact called as $\Delta I = 1/2 $ rule.
 The magnitude of $\epsilon^\prime$ is smaller than $\epsilon$
at least by this factor
and the determination of $\epsilon^\prime$
is one of the main purpose of experiment at FNAL.

\section{Examples of CPT test}
\label{sec:cpttest}

 It would be ideal to extract the values or the restriction
on the various CP, T and/or CPT violating parameters
by observing, for instance,
the time dependence of such a decay mode
as the kaon into two pions in detail.
 But the number of the parameters to be determined
is actually quite limited.
 Another direction of CPT test is to find a relation
implied from CPT invariance
which can be expressed in terms of the quantities
available from the measurement
and check whether such a relation is realized or not.
 A example of the latter is
the relation \cite{Tanner}
\begin{equation}
 \lambda_0
  = {\rm Re}\,\left( \epsilon_0 - \Delta_{\rm CPT} \right)
   - {\rm Re}\,\eta_{+-} + {\rm Re}\,\epsilon^\prime,
  \label{eq:cpt-test1}
\end{equation}
where $\lambda_0$ represents the CPT violation
as we have seen in the previous section.
 Each quantity in the right handed side
can be determined from the observation
of the time dependence of the charged pion decay mode
if we neglect the last term which is smaller
than the others by $\Delta I = 1/2 $ rule.

 The time dependence of the neutral kaon
to the charged pion is described in the following manner.
 Let us call the state $ \psi $ which is initially $K^0$
($\bar{K}^0$) as $\psi_K$ ($\psi_{\bar{K}}$).
 Then
\begin{eqnarray}
 \displaystyle{
   \frac{ \left| \left< \pi^+ \pi^- \left| T
             \right| \psi_K(t) \right>\right|^2 }
        { \left| \left< \pi^+ \pi^- \left| T
             \right| K_S \right>\right|^2 / 2 }
 } &=&
 \displaystyle{
  A_K e^{-\Gamma_S t}
  + 2 \left\{
        C_K e^{-\bar{\Gamma} t} \cos(\Delta m t)
        + S_K e^{-\bar{\Gamma} t} \sin(\Delta m t)
      \right\}
  + B_K e^{-\Gamma_L t}
 }, \nonumber \\
 \displaystyle{
   \frac{ \left| \left< \pi^+ \pi^- \left| T
             \right| \psi_{\bar{K}}(t) \right>\right|^2 }
        { \left| \left< \pi^+ \pi^- \left| T
             \right| K_S \right>\right|^2 / 2 }
 } &=&
 \displaystyle{
  A_{\bar{K}} e^{-\Gamma_S t}
  + 2 \left\{
        C_{\bar{K}} e^{-\bar{\Gamma} t} \cos(\Delta m t)
        + S_{\bar{K}} e^{-\bar{\Gamma} t} \sin(\Delta m t)
      \right\}
  + B_{\bar{K}} e^{-\Gamma_L t}
 }, \nonumber \\
 \label{eq:time-dep}
\end{eqnarray}
where
$ \bar{\Gamma} = (\Gamma_S + \Gamma_L)/2 $
and $\Delta m = m_L = m_S $.
 The various coefficients in (\ref{eq:time-dep}) are given by
\begin{eqnarray}
 A_K &=&
 \displaystyle{
  1 - 2 {\rm Re}\,\left( \epsilon_0 - \Delta_{\rm CPT} \right)
 }, \nonumber \\
 C_K &=&
 \displaystyle{
  \left( 1 - 2\,{\rm Re}\epsilon_0 \right) {\rm Re}\,\eta_{\pi^+\pi^-}
    + 2\,{\rm Im}\Delta_{\rm CPT}\,{\rm Im}\,\eta_{\pi^+\pi^-}
 }, \nonumber \\
 S_K &=&
 \displaystyle{
  \left( 1 - 2\,{\rm Re}\epsilon_0 \right) {\rm Re}\,\eta_{\pi^+\pi^-}
    - 2\,{\rm Im}\Delta_{\rm CPT}\,{\rm Im}\,\eta_{\pi^+\pi^-}
 }, \nonumber \\
 D_K &=&
 \displaystyle{
  \left| \eta_{\pi^+\pi^-} \right|^2
  \left\{
    1 - 2\,{\rm Re}\left( \epsilon_0 + \Delta_{\rm CPT} \right)
  \right\}
 },
 \label{K-coeff}
\end{eqnarray}
for initially $K^0$ state,
while for initially $\bar{K}^0$ state
\begin{eqnarray}
 A_{\bar{K}} &=&
 \displaystyle{
  1 + 2 {\rm Re}\,\left( \epsilon_0 - \Delta_{\rm CPT} \right)
 }, \nonumber \\
 C_{\bar{K}} &=& 
 \displaystyle{ -
  \left( 1 + 2\,{\rm Re}\epsilon_0 \right) {\rm Re}\,\eta_{\pi^+\pi^-}
    + 2\,{\rm Im}\Delta_{\rm CPT}\,{\rm Im}\,\eta_{\pi^+\pi^-}
 }, \nonumber \\
 S_{\bar{K}} &=&
 \displaystyle{ -
  \left( 1 + 2\,{\rm Re}\epsilon_0 \right) {\rm Re}\,\eta_{\pi^+\pi^-}
    - 2\,{\rm Im}\Delta_{\rm CPT}\,{\rm Im}\,\eta_{\pi^+\pi^-}
 }, \nonumber \\
 D_{\bar{K}} &=&
 \displaystyle{
  \left| \eta_{\pi^+\pi^-} \right|^2
  \left\{
    1 + 2\,{\rm Re}\left( \epsilon_0 + \Delta_{\rm CPT} \right)
  \right\}
 }.
 \label{bK-coeff}
\end{eqnarray}
 That the same combination appears
as the coefficients in the same exponentially damping terms in
(\ref{K-coeff}) and (\ref{bK-coeff})
is evident from the reverse of (\ref{eq:eigen})
\begin{eqnarray}
&&
 \displaystyle{
  \left| K^0 \right>
  = \frac{1}{\sqrt{2}}
    \left\{
     \left(
       1 - \left( \epsilon_0 - \Delta_{\rm CPT} \right)
     \right) \left| K_S \right>
     - 
     \left(
       1 - \left( \epsilon_0 + \Delta_{\rm CPT} \right)
     \right) \left| K_L \right>
    \right\}
 }, \nonumber \\
&&
 \displaystyle{
  \left| \bar{K}^0 \right>
  = \frac{1}{\sqrt{2}}
    \left\{
     -
     \left(
       1 + \left( \epsilon_0 - \Delta_{\rm CPT} \right)
     \right) \left| K_S \right>
     - 
     \left(
       1 + \left( \epsilon_0 + \Delta_{\rm CPT} \right)
     \right) \left| K_L \right>
    \right\}
 },
\end{eqnarray}
which hold up to the first order
in $\epsilon_0$ or $\Delta_{\rm CPT}$.
 The last terms are small as they are second order in $\eta_{\pi^+\pi^-}$.
 The other terms are governed by the scale of life time of $K_S$.
 From the data in the short time range, $ \lesssim 4/\Gamma_S $,
the ${\rm Re}\,(\epsilon_0 - \Delta_{\rm CPT}) $ will be
determined whereas
in the range $ 4 \lesssim t/\Gamma_S \lesssim 16 $
the oscillating terms dominate so that they can provide
the nice measurement of ${\rm Re}\,\eta_{\pi^+\pi^-}$.
 Thus combined with (\ref{eq:cpt-test1}) CPT test is available.

 Another example of CPT test is based on 
the Bell-Steinberger relation \cite{Bell}
\begin{eqnarray}
 \displaystyle{
  \frac{d}{dt} \left| \psi(t) \right|^2
 = \sum_n \left| \left< n \left| T
      \right| \psi(t) \right> \right|^2
 },
\end{eqnarray}
which is nothing but the unitarity requirement
has been taken up as a check of CPT invariance. 
 More explicitly from Eqs. (\ref{eq:time}) and (\ref{eq:eigen})
\begin{eqnarray}
 &&
 \displaystyle{
  2 \left(
     {\rm Re} \epsilon_0 - i {\rm Im} \Delta_{\rm CPT}
    \right)
    \left(
     i \Delta M + \bar{\Gamma}
    \right)
  =
  \sum_n A(K_L \rightarrow n)^* A(K_S \rightarrow n)
 },
 \label{eq:red-BS}
\end{eqnarray}
 The total decay width of the neutral kaon
is almost saturated by a few kind of modes; two pions, semileptonic ones
and three pions,
and they will be sufficient to extract the information
on $\epsilon_0$ and $\Delta_{\rm CPT}$ through (\ref{eq:red-BS}).

 Sometimes this relation or its reduced form
have been used for CPT test \cite{CPLEAR,Fermi},
but with some additional assumptions,
neglecting the CPT violation in the decay amplitudes,
and/or neglecting $\Delta S = \Delta Q$ rule violation
in the semileptonic decay mode.
 This is because ( \ref{eq:red-BS}) involves
a large number of CPT violating sources most generally
as the decay amplitudes of the various channels
appear in this relation.
 Actually the limit (\ref{eq:CPLEAR-limit}) due to CPLEAR
\cite{CPLEAR,Tauscher} appears
to have been obtained based on this assumption.
 Thus we must treat more elaborately
that equation such as in Ref. \cite{Lavoura}.
 Or we allow that $\Delta S = \Delta Q $ rule be largely violated
and disentangle it with
the potential CPT violation in the analysis
\cite{HS}.

\section{Conclusion}
\label{sec:conclusion}

 Here we have seen the most fundamental aspect of CPT test
in the neutral kaon system.
 The existence of CPT violation
will also have phenomenological implication,
such as on baryon number production \cite{baryon1,baryon2}.

 A laking feature of possibility of CPT violation
is that there is no explicit examples which leads CPT violation,
as it may be beyond the description by the quantum field theory. 
 For instance, the perturbative part of the Hamiltonian $H_W$
cannot be a Lorentz invariant local operator if CPT violation
is actually involved in it.
 It might be a non-local operator,
which does not admit the usual operator product expansion.
 Then the optimistic estimate of the effect of CPT violation
in Sec. \ref{sec:expectation} will not be valid.
 Thus the explicit demonstration of CPT violation
in some simple model is an important subject
to convince the phenomenologists to pursue
the CPT violation in the real world.

 In the near future,
$\phi $ factory at Frascati may find or improve the bound on
CPT violation,
but we always have to keep in mind
what was assumed before accepting the result to be presented
as was noted in the previous section.

\acknowledgements{
 The author thanks to K. Hagiwara and M. Tanabashi
for the discussion concerning with the present subject.
}

%

\end{document}